\documentclass[twocolumn,showpacs,amsmath,amssymb,prl,aps,superscriptaddress,floatfix]{revtex4-1}

\usepackage{graphicx}
\usepackage{amsmath}
\usepackage{verbatim}
\usepackage{epstopdf}
\usepackage{pifont}

\usepackage[urlcolor=blue, hyperindex, colorlinks, bookmarks=true]{hyperref}

\newcommand{\key}{\mathbf{k}}
\newcommand{\Va}{V_{A}}
\newcommand{\Vda}{V_{{D}_1}}
\newcommand{\Vdn}{V_{{D}_n}}
\newcommand{\Vdi}{V_{D_i}}
\newcommand{\etad}{\eta_{\mathrm{D}}}
\newcommand{\etadi}{\eta_{{D}_i}}
\newcommand{\etaa}{\eta_{\mathrm{A}}}
\newcommand{\Dvec}{\mathbf{D}}
\newcommand{\Da}{D_1}
\newcommand{\Db}{D_2}
\newcommand{\Dc}{D_3}
\newcommand{\Dd}{D_4}
\newcommand{\Di}{D_i}
\newcommand{\Dn}{D_n}
\newcommand{\CR}{\mathrm{CR}}
\newcommand{\ki}{k_i}
\newcommand{\ZX}{\mathrm{CR}}
\newcommand{\Zpihalf}{Z_{90}}
\newcommand{\CNOT}{\mathrm{CNOT}}
\newcommand{\None}{N_{1\%}}
\newcommand{\Noneavg}{\bar{N}_{1\%}}
\newcommand{\pe}{p}
\newcommand{\peavg}{\bar{p}}
\newcommand{\Zm}{Z_{90}^-}
\newcommand{\Xm}{X_{90}^-}
\newcommand{\Vdvec}{\mathbf{V}_\mathrm{D}}
\newcommand{\etaabar}{\tilde{\eta}_{\mathrm{A}}}
\newcommand{\Pdz}{\rho_0}
\newcommand{\Pdo}{\rho_1}
\newcommand{\uo}{\mu_1}
\newcommand{\uz}{\mu_0}
\newcommand{\varz}{\sigma^2_0}
\newcommand{\varo}{\sigma^2_1}
\newcommand{\Pkz}{\rho_{k_i=0}}
\newcommand{\Pko}{\rho_{k_i=1}}
\newcommand{\A}{A}
\newcommand{\ket}[1]{\left\lvert #1 \right\rangle}

\newcommand{\avg}[1]{\left\langle #1 \right\rangle}
\newcommand{\abs}[1]{\left| #1 \right|}
\newcommand{\MHz}{\mathrm{MHz}}

\usepackage{microtype}

\begin{document}
\title{Demonstration of quantum advantage in machine learning} 

\author{Diego~Rist\`e}

\author{Marcus~P.~da~Silva}

\author{Colm~A.~Ryan}

\affiliation{Raytheon BBN Technologies, Cambridge, MA 02138, USA}

\author{Andrew~W.~Cross}

\author{John~A.~Smolin}

\author{Jay~M.~Gambetta}

\author{Jerry~M.~Chow}

\affiliation{IBM T.J. Watson Research Center, Yorktown Heights, NY 10598, USA}

\author{Blake~R.~Johnson}

\affiliation{Raytheon BBN Technologies, Cambridge, MA 02138, USA}

\date{\today}

\pacs{}

\maketitle

\textbf{
The main promise of quantum computing is to efficiently solve certain problems that are prohibitively expensive for a classical computer. 
Most problems with a proven quantum advantage involve the repeated use of a black box, or oracle, whose structure encodes the solution~\cite{Nielsen00}.
One measure of the algorithmic performance is the query complexity~\cite{Cleve01}, i.e., the scaling of the number of oracle calls needed to find the solution with a given probability.
Few-qubit demonstrations of quantum algorithms, such as Deutsch-Jozsa and Grover~\cite{Nielsen00}, have been implemented across diverse physical systems such as nuclear magnetic resonance~\cite{Jones98, Linden98, Chuang98a, Chuang98b}, trapped ions~\cite{Gulde03}, optical systems~\cite{Takeuchi00, Kwiat00}, and superconducting circuits~\cite{DiCarlo09, Yamamoto10, Dewes12}. 
However, at the small scale, these problems can already be solved classically with a few oracle queries, and the attainable quantum advantage is modest~\cite{Yamamoto10, Dewes12}. 
Here we solve an oracle-based problem, known as learning parity with noise~\cite{Angluin88, Blum03}, using a five-qubit superconducting processor. 
Running classical and quantum~\cite{Cross15} algorithms on the same oracle, we observe a large gap in query count in favor of quantum processing. We find that this gap grows by orders of magnitude as a function of the error rates and the problem size. This result demonstrates that, while complex fault-tolerant architectures will be required for universal quantum computing, a quantum advantage already emerges in existing noisy systems.}

The limited size of engineered quantum systems and their extreme susceptibility to noise sources have made it hard so far to establish a clear advantage of quantum over classical computing.  
A promising avenue to highlight this separation is offered by a new family of algorithms designed for machine learning~\cite{Schuld15, Manzano09, Lloyd13, Wiebe14}. 
In this class of problems, artificial intelligence methods are employed to discern patterns in large amounts of data, with little or no knowledge of underlying models. 
A particular learning task, known as binary classification, is to identify an unknown mapping between a set of bits onto $0$ or $1$.   
An example of binary classification is identifying a hidden parity function~\cite{Angluin88, Blum03}, defined by the unknown bit-string $\key$, which computes $f(\Dvec,\key) = \Dvec\cdot \key \mod 2$ on a register of $n$ data bits $\Dvec = \{\Da, \Db ..., \Dn\}$ (Fig.~1a). 
The result, i.e., 0 (1) for even (odd) parity, is mapped onto the state of an additional bit $\A$. The learner has access to the output register of an \emph{example oracle} circuit that implements $f$ on random input states, on which he/she has no control. Repeated queries of the oracle allow the learner to reconstruct $\key$. 
However, any physical implementation suffers from errors, both in the oracle execution itself and in readout of the register. In the presence of errors, the problem becomes hard. Assuming that every bit introduces an equal error probability, the best known algorithms have a number of queries growing as $\mathcal{O}(n)$ and runtime growing almost exponentially with $n$~\cite{Angluin88, Blum03,Lyubashevsky05}. In view of the classical hardness of learning parity with noise (LPN), parity functions have been suggested as keys for secure and computationally easy authentication~\cite{Hopper01, Pietrzak12}. 

The picture is different when the algorithm can process quantum superpositions of input states, i.e., when the oracle is implemented by a quantum circuit. In this case, applying a coherent operation on all qubits after an oracle query ideally creates the entangled state 
\begin{equation}
\label{eq:LPNq}
(\ket{0_A0_\textbf{D}^n} + \ket{1_A\key_\Dvec})/\sqrt{2}.
\end{equation}
In particular, when $\A$ is measured to be in $\ket{1}$, $\ket{\textbf{D}}$ will be projected onto $\ket{\key}$. 
With constant error per qubit, learning from a quantum oracle requires a number of queries that scales as $\mathcal{O}(\log n)$, and has a total runtime that scales as $\mathcal{O}(n)$~\cite{Cross15}. This gives the quantum algorithm an exponential advantage in query complexity and a super-polynomial advantage in runtime.

In this work, we implement a LPN problem in a superconducting quantum circuit using up to five qubits, realizing the experiment proposed in Ref.~\onlinecite{Cross15}. We construct a parity function with bit-string $\key$ using a series of $\CNOT$ gates between the ancilla and the data qubits (Fig.~1b). We then present two classes of learners for $\key$ and compare their performance. The first class simply measures the output qubits in the computational basis and analyzes the results. 
The measurement collapses the state into a random $\{\Dvec, f(\Dvec, \key)\}$ basis state, reproducing an example oracle of the classical LPN problem.  
The second class performs some quantum computation (coherent operations), followed by classical analysis, to infer the solution. 
We show that the quantum approach outperforms the classical one in the number of queries required to reach a target error threshold, and that it is largely robust to noise added to the output qubit register. 

\begin{figure}
\includegraphics[width=\columnwidth]{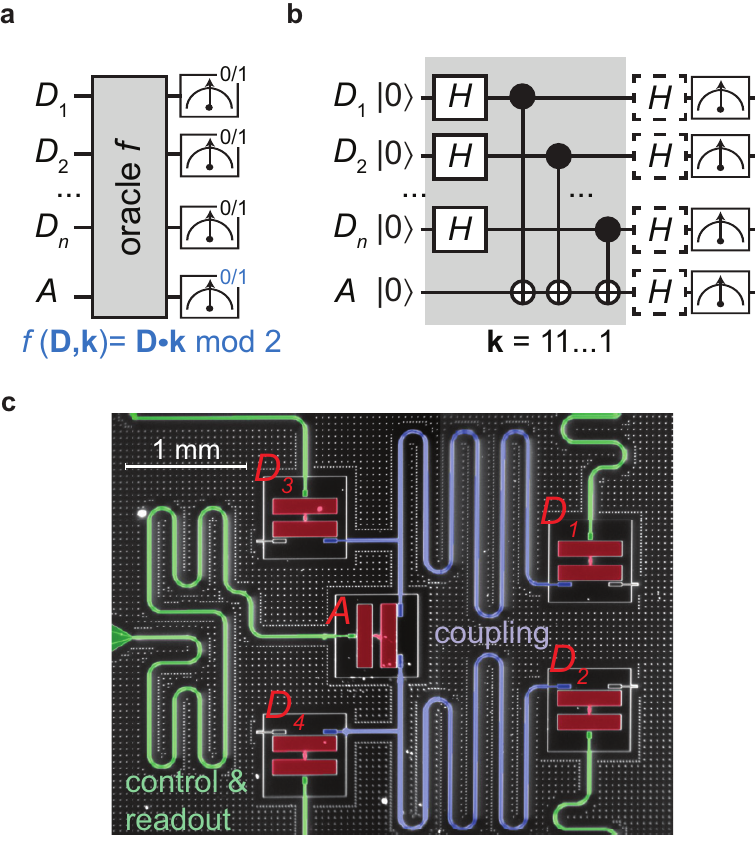}
\caption{\textbf{Implementation of a parity function in a superconducting circuit.} (\textbf{a}) Conceptual diagram of parity learning. The (classical or quantum) oracle $f$ ideally maps the parity of a subset of $n$ data bits (or qubits), defined by the bit string $\key$, into bit $A$. Repeated queries of the oracle allow the reconstruction of $\key$ by reading the output register. (\textbf{b}) Gate sequence implementing a quantum parity oracle with $\key=11...1$. Random examples are generated by preparing the data qubits $\{\Da, ..., \Dn\}$ in a uniform superposition. Vertical lines indicate $\CNOT$ gates between each $\Di$ (control) and the ancilla qubit $\A$ (target). Quantum learning differs from classical learning only by the addition of single-qubit gates (dashed boxes) applied before measurement (see also Extended Data Fig.~1). 
(\textbf{c}) Optical image of the superconducting quantum processor (qubits in red). $\A$ is coupled to each $\Di$ by means of two bus resonators (blue). Each qubit is also coupled to a dedicated resonator for control and readout (green)~\cite{Corcoles15}.}
\end{figure}
The quantum device used in our experiment consists of five superconducting transmon qubits, $\A, \Da, ..., \Dd$, and seven microwave resonators (Fig.~1c). 
Five of the resonators are used for individual control and readout of the qubits, to which they are dispersively coupled~\cite{Blais04}. The center qubit $\A$ plays the role of the result and is coupled to the data register $\{\Di\}$ via the remaining two resonators. This coupling allows the implementation of cross-resonance ($\CR$) gates~\cite{Rigetti10} between $\A$ (used as control qubit) and each $\Di$ (target), constituting the primitive two-qubit operation for the circuit in Fig.~1b (full gate decomposition in Extended Data Fig.~1). Each qubit state is read out by probing the dedicated resonator with a near-resonant microwave pulse. The output signals are then demodulated and integrated at room temperature to produce the homodyne voltages $\{\Vda, ... \Vdn, \Va\}$ (see Extended Data Fig.~2 for the detailed experimental setup). 

To implement a uniform random example oracle for a particular $\key$, we first prepare the data qubits in a uniform superposition (Fig.~1b). Preparing such a state ensures that all parity examples are produced with equal probability and is also key in generating a quantum advantage. We then implement the oracle as a series of CNOT gates, each having the same target qubit $\A$ and a different control qubit $\Di$ for each $\ki=1$. Finally, the state of all qubits is read out (with the optional insertion of Hadamard gates, see discussion below).   
The oracle mapping to the device is limited by imperfections in the two-qubit gates, with average fidelities $88 - 94\%$, characterized by randomized benchmarking~\cite{Magesan12} (see Extended Data Table~1). Readout errors in the register $\etadi$, defined as the average probability of assigning a qubit to the wrong state, are limited to $20-40\%$ by the onset of inter-qubit crosstalk at higher measurement power (Extended Data Fig.~3). A Josephson parametric amplifier~\cite{Hatridge11} in front of the amplification chain of $\A$ suppresses its low-power readout error to $\etaa = 5\%$.

Having implemented parity functions with quantum hardware, we now proceed to interrogate an oracle $N$ times and assess our capability to learn the corresponding $\key$. We start with oracles with register size $n=2$, involving $\Da, \Db$, and $\A$. We consider two classes of learning strategies, classical (C) and quantum (Q). In C, we perform a projective measurement of all qubits right after execution of the oracle. 
This operation destroys any coherence in the oracle output state, thus making any analysis of the result classical. 
The measured homodyne voltages $\{\Vda, ... \Vdn, \Va\}$ are converted into binary outcomes, using a calibrated set of thresholds (see Methods). 
Thus, for every query, we obtain a binary string $\{a, d_1, d_2\}$, where each bit is $0$ ($1$) for the corresponding qubit detected in $\ket{0}$ ($\ket{1}$). Ideally, $a$ is the linear combination of $d_1, d_2$ expressed by the string $\key$ (Fig.~1a). However, both the gates comprising the oracle and qubit readout are prone to errors (see Extended Data Table~1). To find the $\key$ that is most likely to have produced our observations, at each query $m$ we compute the expected $\tilde{a}_{\key,m}$ for the measured $d_{1,m}, d_{2,m}$ and the $4$ possible values of $\key$. We then select the $\key$ which minimizes the distance to the measured results $a_1, ..., a_N$ of $N$ queries, i.e., $\sum^N_m{\abs{\tilde{a}_q - a_{i,\key}}}$~\cite{Angluin88}. In the case of a tie, $\key$ is randomly chosen among those producing the minimum distance. As expected, the error probability $p$ of obtaining the correct answer decreases with $N$ (Fig.~2a). Interestingly, the difficulty of the problem depends on $\key$ and increases with the number of $\ki=1$. This can be intuitively understood as needing to establish a higher correlation between data qubits when the weight of $\key$ increases. 

\begin{figure}
\includegraphics[width=\columnwidth]{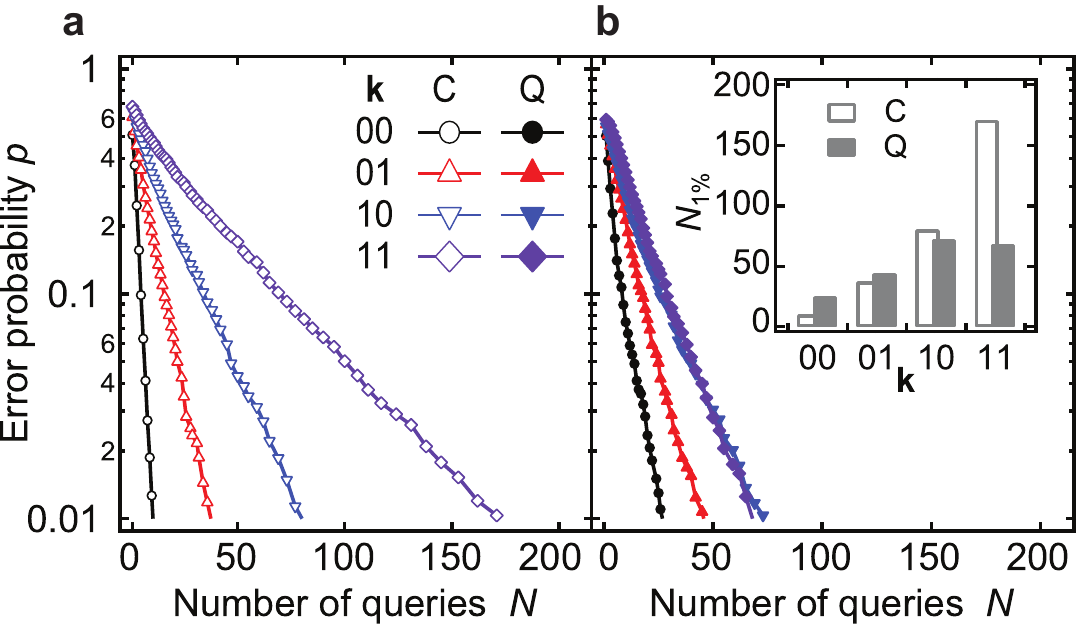}
\caption{\textbf{Error probability $\pe$ to identify a 2-bit oracle $\key$ as a function of the number of queries $N$.} For both classical (\textbf{a}) and quantum (\textbf{b}) learners, one of the four oracles $\key$ is applied, followed by the simultaneous measurement of all qubits. Hadamard gates are applied prior to measurement in the quantum case (Fig.~1b). See text for a description of the solvers in the two scenarios. Inset: number of queries $\None(\key)$ required to reach $1\%$ error for the classical (empty bars) and quantum (solid) solver.}
\end{figure}

In our second approach (Q), while the oracle is left untouched, we apply local operations (Hadamard gates) to all qubits before measuring. Remarkably, this simple operation completely changes the statistics of the measurement results and the learning procedure. 
We now use the fact that the state of the data qubits is entangled with the result $\A$ (see Eq.~\ref{eq:LPNq}). Whenever $\A$ is measured to be in $\ket{1}$, the data register will ideally be projected onto the solution, $\ket{\Da, \Db} = \ket{k_1, k_2}$. We therefore digitize and postselect our results on the $\approx50\%$ outcomes where $a = 1$ and perform a bit-wise majority vote on $\{d_1, d_2\}_{1...\tilde{N}}$. 
Despite every individual query being subject to errors, the majority vote is effective in determining $\key$ (Fig.~2b). We assess the performance of the two solvers by comparing the number of queries $\None$ required to reach $\pe=0.01$ (Fig.~2c). 
Whereas Q performs comparably or worse than C for $\key = 00, 01$ or $10$, Q requires less than half as many queries as C for the hardest oracle, $\key=11$. We note that, while these results are specific to our the lowest oracle and readout errors we can achieve (see Extended Data Table~1), a systematic advantage of quantum over classical learning will become clear in the following. 

So far we have adhered to a literal implementation of the classical LPN problem, where each output can only be either 0 or 1.  
However, the actual measurement results are the continuous homodyne voltages $\{\Vda, ... \Vdn, \Va\}$, each having mean and variance determined by the probed qubit state and by the measurement efficiency, respectively~\cite{Blais04}. These additional resources can be exploited to improve the learner's capabilities as follows. A more effective strategy for C uses Bayesian estimation to calculate the probability of any possible $\key$ for the measured output voltages, and select the most probable (see Methods). This approach is expensive in classical processing time (scaling exponentially with $n$), but drastically reduces the error probability $\peavg$, averaged over all $\key$, at any $N$ (Fig.~3 and Extended Data Fig.~4). To improve Q, we still postselect each oracle query on digital $a=1$, but average all instances of $\{\Vdi\}$, and digitize the averages $\{\avg{\Vdi}\}$ instead of each observation (see Methods). For each $\Di$, the majority vote between $\approx N/2$ inaccurate observations is then replaced by a single vote with high accuracy. Using the analog results, not only does Q retain an advantage over C (smaller $\pe$ for given $N$), but it does so without introducing an overhead in classical processing.    

\begin{figure}
\includegraphics[width=\columnwidth]{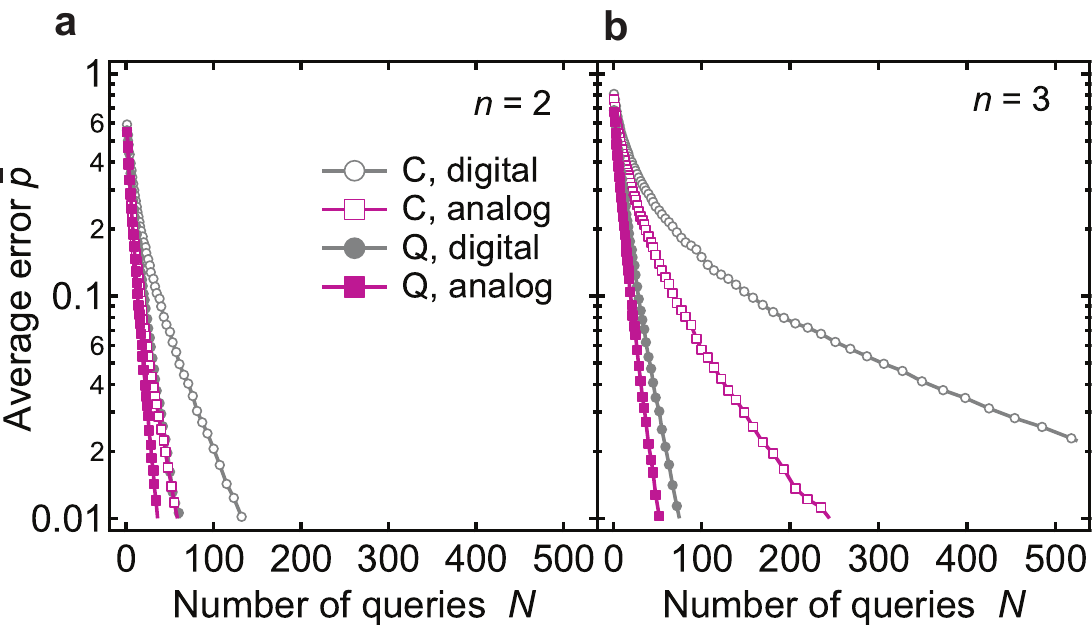}
\caption{\textbf{Learning error probability $\peavg$ averaged over all the $n$-bit oracles $\key$, for different $n$ and solvers}. (\textbf{a}) $n=2$,   (\textbf{b}) $n=3$. Making use of the analog measurements results $\{\Vda, ... \Vdn, \Va\}$ (squares) improves over the digital solvers in Fig.~2 (circles) for both classical (empty symbols) and quantum (solid symbols) learning.  
The analog solver in Q proves to be the most efficient solution. Moreover, the gap between Q and C grows with $n$. The same dataset is used in Figs.~2 and 3, with $\Dc$ ignored in the analysis for $n=2$. See Extended Data Fig.~4 for the $\pe (N)$ corresponding to each $3$-bit $\key$.}
\end{figure}

The superiority of Q over C becomes even more evident when the oracle size $n$ grows from 2 to 3 data qubits (Fig.~3b). Whereas Q solutions are marginally affected, the best C solver demands almost an order of magnitude higher $N$ to achieve a target error. Maximizing the resources available in our quantum hardware, we observe an even larger gap for oracles with $n=4$ (Extended Data Fig.~5), suggesting a continued increase of quantum advantage with the problem size.

As predicted, quantum parity learning surpasses classical learning in the presence of noise. To investigate the impact of noise on learning, we introduce additional readout error on either $\A$ or on all $\Di$. This can be easily done by tuning the amplitude of the readout pulses, effectively decreasing the signal-to-noise ratio~\cite{Vijay11}. When the ancilla assignment error probability $\etaa$ grows (Fig.~4a), the number of queries $\Noneavg$ (the average of $\None$ over all $\key$) required by the C solver increases by up to 2 orders of magnitude in the measured range (see also Extended Data Fig.~6).  
Conversely, using Q, $\Noneavg$ only changes by a factor of $\sim3$. Key to this performance gap is the optimization of the digitization threshold for $\{\avg{\Vdi}\}$ at each value of $\etaa$ (see Methods). 
When $\etaa$ is increased, an interesting question is whether postselection on $\Va$ remains always beneficial. In fact, for $\etaa>0.25$, it becomes more convenient to ignore $\Va$ and use the totality of the queries ($\mathrm{Q}'$ in Fig.~4a).

\begin{figure}
\includegraphics[width=\columnwidth]{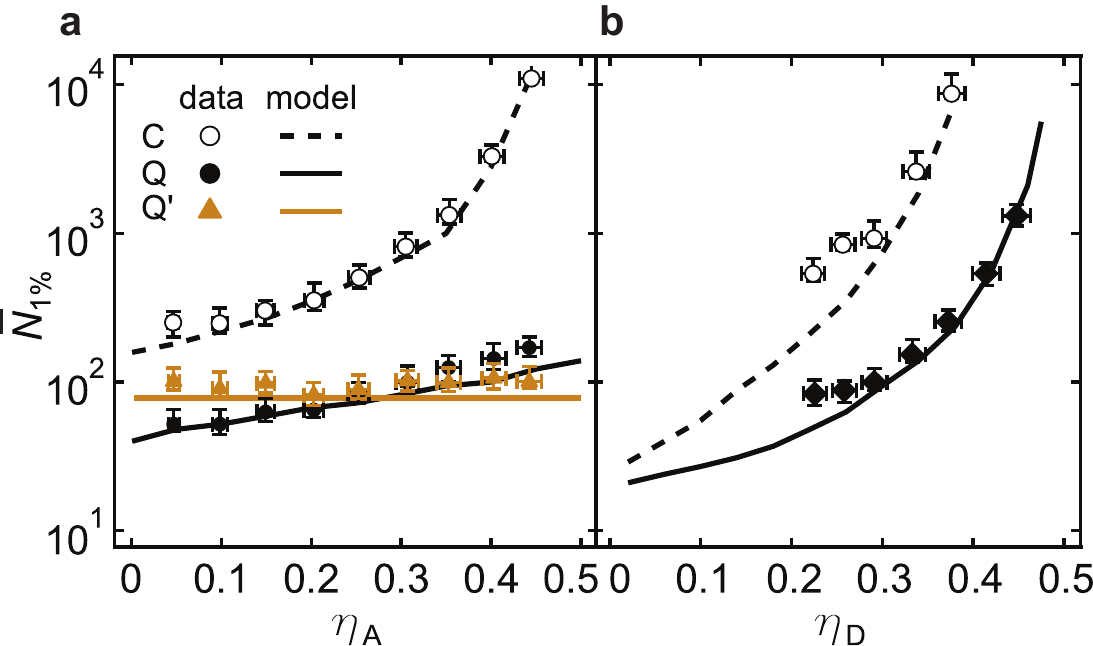}  
\caption{\textbf{Robustness of quantum parity learning to noise.} Number of queries $\Noneavg$ for $\peavg = 0.01$ for variable readout error $\eta$ of ancilla (\textbf{a}) or data (\textbf{b}) qubits, with $n=3$. $\eta$ is tuned by setting the readout power of the corresponding qubit(s). Empty (solid) circles correspond to the analog  C (Q) solver. (\textbf{a}), $\Noneavg$ diverges for $\etaa \to 0.5$ for C, while it stays limited for Q. When $\etaa \gtrsim 0.25$, it is preferable to ignore $\Va$ altogether ($\mathrm{Q}'$, triangles). (\textbf{b}) Whereas both C and Q are severely affected by a noisy data register, Q remains superior and the performance gap increases with $\etad$. See Methods for an explanation of the error bars.}
\end{figure}

Similarly, we step the readout error of the data qubits, with average $\etad$, while setting $\etaa$ to the minimum. Not only does Q outperform C at every step, but the gap widens with increasing $\etad$. 

A numerical model including the measured $\etaa, \etad$, qubit decoherence, and gate errors modeled as depolarization noise (Extended Data Table 1) is in very good agreement with the measured $\None$ at all $\etaa, \etad$. This model allows us to extrapolate $\None$ to the extreme cases of zero and maximum noise. Obviously, when $\etad  = 0.5$, readout of the data register contains no information, and $\None$ consequently diverges. On the other hand a random ancilla result ($\etaa = 0.5$) does not prevent a quantum learner from obtaining $\key$. In this limit, the predicted factor of $\sim2$ in $\Noneavg$ between Q and $\mathrm{Q}'$ can be intuitively understood as Q indiscriminately discards half of the queries, while $\mathrm{Q}'$ uses all of them. (See Supplementary Material for theoretical bounds on the scaling of $\Noneavg$ for different solvers.)

In conclusion, we have implemented a learning parity with noise algorithm in a quantum setting. We have demonstrated a superior performance of quantum learning compared to its classical counterpart, where the performance gap increases with added noise in the query outcomes.   
A quantum learner, with the ability of physically manipulating the output of a quantum oracle, is expected to find the hidden key with a logarithmic number of queries and linear runtime as function of the problem size, whereas a passive classical observer would require a linear number of queries and nearly exponential runtime. We have shown that the difference in classical and quantum queries required for a target error rate grows with the oracle size in the experimentally accessible range, and that quantum learning is much more robust to noise. 
We expect that future experiments with increased oracle size will further demarcate a quantum advantage, in support of the predicted asymptotic behavior.

\section{Methods}

\textbf{Pulse calibration.}  
Single- and two-qubit pulses are calibrated by an automated routine, executed periodically during the experiments. For each qubit, first the transition frequency is calibrated with Ramsey experiments. 
Second, $\pi$ and $\pi/2$ pulse amplitudes are calibrated using a phase estimation protocol~\cite{Kimmel15}. The pulse amplitudes, modulating a carrier through an I/Q mixer (Extended Data Fig.~2) are adjusted at every iteration of the protocol until the desired accuracy or signal-to-noise limit is reached. 
Pulses have a Gaussian envelope in the main quadrature and derivative-of-Gaussian in the other, with DRAG parameter~\cite{Motzoi09} calibrated beforehand using a sequence amplifying phase errors~\cite{Lucero10}.  
$\ZX$ gates are calibrated in a two-step procedure, determining first the optimum duration and then the optimum phase for a $ZX_{90}$ unitary.  

\textbf{Experimental setup.} 
A detailed schematic of the experimental setup is illustrated in Extended Data Fig.~2. For each qubit, signals for readout and control are delivered to the corresponding resonator through an individual line through the dilution refrigerator. For an efficient use of resources, we apply frequency division multiplexing~\cite{Jerger12} to generate the five measurement tones by sideband modulation of three microwave sources. Moreover, the same pair of BBN APS (custom arbitrary waveform generators) channels produce the readout pulses for $\{\Da, \Db\}$, and another one for $\{\Dc, \Dd\}$. Similarly, the output signals are pairwise combined at base temperature, limiting the number of HEMTs and digitizer channels to three. The attenuation on the input lines, distributed at different temperature stages, is a compromise between suppression of thermal noise impinging on the resonators (affecting qubit coherence) and the input power required for $\ZX$ gates. 

\textbf{Gate sequence.} 
$\CNOT$ gates can be decomposed in terms of $\ZX$ gates using the relation $\CNOT_{12} = (\Zm \otimes \Xm)\, \ZX_{12}$~\cite{Chow14}. Moreover, the role of control and target qubits are swapped, using $\CNOT_{12} = (H_1 \otimes H_2)\, \CNOT_{21} (H_1 \otimes H_2)$. The first of these $H$ gates is absorbed into state preparation for the LPN sequence (Figs.~1a and Extended Data Fig.~1). Similarly, when two $\CNOT$s are executed back to back, two consecutive $H$ gates on $\A$ are canceled out. In order to maintain the oracle identical in C and Q, we do not compile the $H$ gates in the $\CNOT$s with those applied before measurement in Q. 

\textbf{Data analysis.} 
For each set of $\{\key, \etaa, \etad\}$, solver type, and register size $n$, we measure the result of $10,000$ oracle queries. 
Each set is accompanied by $n+2$ calibration points (averaged $10,000$ times), providing the distributions of $\Va, \Vda, ..., \Vdn$ for the collective ground state and for single-qubit excitations ($n$ data and 1 ancilla qubit). These distributions are then used to determine the optimum digitization threshold (for digital solvers) or as input to the Bayesian estimate in C.  
To obtain $\pe (N)$, we resample the full data set with $1000-4000$ random subsets of each size $N$. 
 
Error bars are obtained by first computing the credible intervals for $\pe$ at each set
$\{N,\key, \etaa, \etad\}$. These intervals are computed with
Jeffreys beta distribution prior
$\mathrm{Beta}(\frac{1}{2},\frac{1}{2})$ for Bernoulli
trials, with a credible level of
$100\%-(100\%-95\%)/8\approx99.36\%$. This ensures that, under a union
bound, the average of estimates for $8$ different keys is inside the credible interval
with a probability of at least $95\%$. We then perform antitonic
regression on the upper and lower bounds of the credible
intervals to ensure monotonicity as function of $N$, and find the
intercept to $\pe=0.01$ for each $\key$.  The bounds on the value
$\overline{N}_{1\%}$ averaged over the keys is computed by interval
arithmetic on the credible intervals of $N_{1\%}$ for each $\key$.

\textbf{Classical solver with Bayesian estimate.} An improved
classical solver for the LPN problem can be constructed when the
oracle provides an analog output. Under the assumption of Gaussian
distributions for each possible bit value, this improved solver
corresponds to a Bayesian estimate of the key after a series of
observations of the data and ancilla bits. More formally, taking a
uniform prior distribution for all binary strings produced by the
oracle, one computes the (unnormalized) posterior $p(D_i)$ distribution for each data
bit $D_i$ the output of the oracle,  
\begin{align*}
p(D_i=b|\Vdi) = 
\frac{1}{2} \exp\left[-\frac{(\Vdi-b)^2}{2\sigma_{i}^2}\right]
\end{align*}
The (unnormalized) posterior distribution $p_m(\key|\Vdvec,V_{A})$ for the key
$\key$ after the $m$th query, on the other hand, is given by
\begin{align*}
p_m(\key|\Vdvec,\Va) =  
\exp\left[-\frac{(\Va-\Dvec\cdot\key)^2}{2\sigma_A^2}\right] p(\Dvec|\Vdvec)p_{m-1}(\key),
\end{align*}
where $p_0(\key)$ is the prior distribution for each key. Here and above,  $\{\Vda, ... \Vdn, \Va\}$ are 
rescaled to have mean $0$ and $1$ for the corresponding qubit in $\ket{0}$ and $\ket{1}$, respectively. 
Iterating this procedure (while updating $p(\key)$ at each
iteration), and then choosing the most probable key $\key_{\text{Bayes}} = \arg
\max_{\key} p(\key)$, one obtains an estimate for the key.

\textbf{Analog quantum solver with postselection on $\A$.}
While postselection on $\A$ is performed equally on both digital (Fig.~2) and analog (Figs.~3-4) Q solvers, in the analog case all postselected $\{\Vdi\}$ are averaged together. Finally, the results $\{\avg{\Vdi}\}$ are digitized to determine the most likely $\key$. The choice of digitization threshold for each $\Di$ depends on: a) the readout voltage distributions  $\Pdz$ and $\Pdo$ for the two basis states, each characterized by a mean $\mu$ and a variance $\sigma^2$; b) $\etaa$. Ideally ($\etaa=0$ and perfect oracle), the distribution of each query output $\Vdi$ matches $\Pdz$ ($\Pdo$) for $\ki=0\, (1)$. When $\etaa >0$, the distribution for $\ki =1$ becomes the mixture $\Pko = \etaa\Pdz + (1-\etaa)\Pdo$. This mixture has mean $(1-\etaa)\uo + \etaa \uz$ and variance $(1-\etaa) \varo+ \etaa \varz - 2\etaa(1-\etaa)\uz\uo$. Instead, $\Pkz = \Pdz$ independently of $\etaa$. We approximate the expected distribution of the mean $\avg{\Vdi}$ with a Gaussian having average and variance obtained from $\Pkz (\Pko)$ for $\ki=0\,(1)$. Finally, we choose the digitization threshold for $\Vdi$ which maximally discriminates these two Gaussian distributions. We note that the number of queries scales the variance of both distributions equally and therefore does not affect the optimum threshold. Furthermore, this calibration protocol is independent of the oracle (see Extended Data Fig.~7). 

\textbf{Analog quantum solver without postselection.}
The analysis without ancilla ($\mathrm{Q}'$) closely follows the steps outlined in the last paragraph. For the purpose of extracting the optimum digitization thresholds, we consider $\etaa=0.5$ in the expressions above. This corresponds to an equal mixture of $\Pdz$ and $\Pdo$ when $\ki=1$.

\textbf{Bounds on performance of the analog quantum solvers.} 
Here we demonstrate how the bounds from Ref.~\onlinecite{Cross15} can be easily adapted to
the case where the solver uses analog voltage measurements. We
consider both the case where experiments are postselected based on
the digitized value of the ancilla (referred below as postselected soft
averaging), and the case where the ancilla is ignored altogether. We consider
different error rate for the ancilla and the data qubits.

\textit{Postselected soft averaging.}
In order to generalize the analysis in Ref.~\onlinecite{Cross15} to the
postselected soft averaging case, we now need to take two types of
data errors into account: depolarizing errors (our crude model for
oracle errors), and measurement error (additive Gaussian noise).

First, postselection works identically to
Ref.~\onlinecite{Cross15}, since we treat the ancilla digitally. We note
that, in this analysis, the ancilla error rate combines oracle
errors and readout errors. Given $n$ queries, $n'$ are
postselected according to the ancilla value $\Va$, and $s$ of this
postselections are correct. Although $s$ is unknown in an experiment,
we condition our results on $s$ being typical (i.e., we only consider
the values of $s$ that occur with probability higher than $1-\epsilon$
for some small $\epsilon$.).

For the correct postselections, we have two possible voltage
distributions for each $\Di$, depending on whether the outcome is 0 or 1. The
distribution of the outcomes will depend on whether we have one of the
correct postselections, and on the value of $i$th key bit $\ki$. 
If $\ki=0$, the conditional voltage distributions, depending on
whether we postselected correctly ($\checkmark$) or not (\ding{55}),
are
\begin{align*}
\rho_{i|0}^\checkmark &\sim \mathcal{N}(\eta_d s,s\sigma^2),\\
\rho_{i|0}^\text{\ding{55}} &\sim \mathcal{N}[\eta_d(N'-s),(N'-s)\sigma^2],
\end{align*}
respectively, with $\mathcal{N}(\mu,\sigma^2)$ the normal distribution
with mean $\mu$ and variance $\sigma^2$. Therefore,
the overall distribution is 
\begin{align*}
\rho_{i|0} & \sim \mathcal{N}(\eta_d N',N'\sigma^2).
\end{align*}
If the true bit value is 1, we have
\begin{align*}
\rho_{i|1}^\checkmark &\sim \mathcal{N}((1-\etad)s,s\sigma^2),\\
\rho_{i|1}^\text{\ding{55}} &\sim \mathcal{N}[\etad(N'-s),(N'-s)\sigma^2],
\end{align*}
and therefore
\begin{align*}
\rho_{i|1} &\sim \mathcal{N}[(1-\etad)s+\etad(N'-s),n'\sigma^2].
\end{align*}
Now we must compute the optimal voltage threshold which determine the digital
decision at each of the data qubits. If we define
\begin{align*}
\mu_{i|j} &= \mathbb{E}[\rho_{i|j}],
\end{align*}
the threshold we must choose is
\begin{align*}
T &= \frac{1}{2}\mu_{i|0}+\frac{1}{2}\mu_{i|1} \\
&= \frac{s(1-2\etad)+2\etad N'}{2}.
\end{align*}
The complication is that this is conditioned on $s$, but we will deal
with that later, as the dependence on $s$ also comes from the
distribution of outcomes (not just the threshold). In the following we
assume the value of $s$ to be typical (i.e., $s$ is contained in the
region around the median excluding the distribution tails that add up
to at most some small $\epsilon$). Under this assumption, we require
that $\mu_{i|0} \le T \le \mu_{i|1}$.

The probability of having the right answer at a particular bit is the
probability that the averaged voltage is on the correct side of the
threshold (above or below). If the true value of the bit is 0, i.e.,
if $\ki=0$, given the threshold, we can compute
\begin{align*}
\Pr(\rho_i\le T|s,\ki=0) 
& = \Phi\left(\frac{T-\mu_{i|0}}{N'\sigma^2}\right) \\
& = 1-Q\left(\frac{T-\mu_{i|0}}{N'\sigma^2}\right),
\end{align*}
where $\Phi$ is the cumulative distribution function for a normal distribution,
and $Q$ is the tail probability for the normal distribution.
We can place a lower bound on $\Pr(M_j\le T|s,\ki=0)$ with an upper bound on
$Q$. Note that, for the range of interest, the argument of $Q$ is
always positive, so we can use the bound
\begin{align*}
Q(x)  
&< \frac{1}{2}\exp\left(\frac{x^2}{2}\right), \quad x > 0
\end{align*}
and therefore
\begin{align*}
\Pr(\rho_i\le T|s) &\ge 
1 - \frac{1}{2}\exp\left[-\left(\frac{T-\mu_{i|0}}{\sqrt{2 N'}\sigma}\right)^2\right],
\end{align*}
which is nearly what we want---we must now address the dependence on $s$.
One way to restrict the analysis to typical $s$ is to require that,
for $\etaabar=\max\{\etaa,1-\etaa\}$,the probability
$$
\Pr(|s-\mu_s|<\delta' \mu_s) > 1 - 2 \exp\left( - \frac{\delta'^2\etaabar N'}{3}\right)
$$
is exponentially close to 1. This choice of $\etaabar$ requires knowledge
of the error rates in the ancilla so that, for example, one knows to
postselect on 0 instead of 1 if $\etaa>0.5$. 

In order to pick a lower bound valid for all typical thresholds and
means, we choose the smallest $|T-\mu_{i|0}|$ by choosing $T$ and
$\mu_{j|0}$ independently from the typical sets. This leads to
$$
T-\mu_{i|0} > N' (1 - \delta') \left(\frac{1}{2} -  \etad\right) \etaabar 
$$
and thus,
$$
\Pr(\rho_i\le T|s) \ge 
1 - \frac{1}{2}\exp\left[-\frac{N'(1 - \delta')^2 \left(\frac{1}{2} -  \etad\right)^2 \etaabar^2}{2\sigma^2}\right]
$$
so that, by the union bound,
$$
\Pr(\tilde{a}\not=a|s) 
< \frac{n}{2}\exp\left[-\frac{N'(1 - \delta')^2 \left(\frac{1}{2} -  \eta_d\right)^2 \etaabar^2}{2\sigma^2}\right]
$$
and therefore the lower bound on the number of queries is
$$
N' > \frac{2\sigma^2}{(1 - \delta')^2 \left(\frac{1}{2} -  \eta_d\right)^2 \etaabar^2}\ln\frac{n}{2\delta}.
$$

If $\ki=1$, we take a similar approach, but the lower bound on the
distance between the threshold and the mean is smaller, leading to
$$
N' > \frac{2\sigma^2}{(1 - 3\delta')^2 \left(\frac{1}{2} -  \eta_d\right)^2 \etaabar^2}\ln\frac{n}{2\delta},
$$
so clearly this is the worst case for $\ki$. 

If we want to bound $N$ instead of $N'$, we just remember that there
is a $50\%$ chance of collapsing into the informative branch of the
state, and using the same typicality argument as before, we have
$$
N > \frac{4\sigma^2}{(1 - \delta'')^2(1 - 3\delta')^2 \left(\frac{1}{2} -  \eta_d\right)^2 \etaabar^2}\ln\frac{n}{2\delta},
$$
where $\delta''$ measures how far from the mean $k$ is, with a corresponding Chernoff bound.

\emph{Analysis without postselection.}
The analysis is equivalent to the postselected case, but with
$\eta_a=\frac{1}{2}$ and $N'=N$, since we keep all experiments and
have a $50\%$ chance of collapsing the state in the informative branch.
All of this leads to
$$
N > \frac{8\sigma^2}{(1 - 3\delta')^2(1/2-\etad)^2}\ln\frac{n}{2\delta}.
$$
We now see that depending of choices of $\delta'$ and $\delta''$,
postselection may or may not lead to better bounds, but the
asymptotic scaling is the same.

\textbf{Complexity of digital classical solvers.}
Angluin and Laird~\cite{Angluin88} showed that learning with classification
noise requires $O(n)$ queries as long as the classification error rate 
is below $\frac{1}{2}$, and propose an algorithm (\emph{disagreement minimization})
 that corresponds to solving an NP-complete problem. According to the exponential time
hypothesis, it is widely believe that NP-complete problems can only be
solved in exponential time. Note that, while the classification rate is nominally
$\etaa$ in our experiment, all errors (including $\etad$ and gate infidelities) 
can be combined onto an effective, $\key$-dependent, single error rate. 

Blum, Kalai, and Wasserman~\cite{Blum03} devised a sub-exponential
time algorithm for learning with classification errors, as long as the
classification error rate is below
$\frac{1}{2}-\frac{1}{2^{n^\delta}}$ for $\delta<1$, at the cost of
increasing the query complexity to slightly sub-exponential scaling with $n$.

Later, Lyubashevsky~\cite{Lyubashevsky05} devised another sligthly
sub-exponential time algorithm for learning with classification
errors, as long as the classification error rate is below
$\frac{1}{2}-\frac{1}{2^{(\log n)^\delta}}$ for $\delta<1$, but
bringing down the query complexity to $n^{1+\epsilon}$ for
$\epsilon>0$.

Note that the gains over exponential time scaling for these two
algorithms are rather small -- a reduction from $O(2^n)$ to
$O(2^{\frac{n}{\log n}})$ and $O(2^{\frac{n}{\log\log n}})$, respectively.

For $n=3$, the Blum-Kalai-Wasserman algorithm can only tolerate less
than $\frac{3}{8}\approx0.375$ classification error rate, while the
Lyubashevsky algorithm can only tolerate less than
$\frac{1}{2}-\frac{1}{2^{\log 3}}\approx0.033$ classification error
rate. Lyubashevsky's algorithm does not apply to any of the
experiments discussed here because our classification error rates are
too high. The Blum-Kalai-Wasserman algorithm only applies to some of
the experiments discussed here, so for the sake of fair comparison
across all error rates, we use Angluin and Laird's disagreement minimization.

\section{Acknowledgments}
\begin{acknowledgments}
We thank George A. Keefe and Mary B. Rothwell for device fabrication,  T.~Ohki for technical assistance, H.~Krovi for discussions, and I.~Siddiqi for providing the Josephson parametric amplifier. This research was funded by the Office of the Director of National Intelligence (ODNI), Intelligence Advanced Research Projects Activity (IARPA), through the Army Research Office contract no. W911NF-10-1-0324. All statements of fact, opinion or conclusions contained herein are those of the authors and should not be construed as representing the official views or policies of IARPA, the ODNI, or the U.S. Government. 
\end{acknowledgments}

\section{Contributions}
C.A.R. and B.R.J. developed the BBN APS and the data acquisition software, D.R. carried out the experiment, D.R., M.P.S. and B.R.J. performed the data analysis, M.P.S. implemented the solvers and developed the theoretical models, D.R. and M.P.S. wrote the manuscript with comments from the other authors, A.W.C. and J.A.S. contributed to the initial design of the experiment, B.R.J., J.M.C. and J.M.G. supervised the project. 

\section{Author information}
\noindent
The authors declare no competing financial interests. \\
Correspondence: driste@bbn.com or bjohnson@bbn.com.

\renewcommand{\figurename}{Extended Data Fig.}
\setcounter{figure}{0}    

\newpage

\begin{figure*}
\includegraphics[width=0.8\columnwidth]{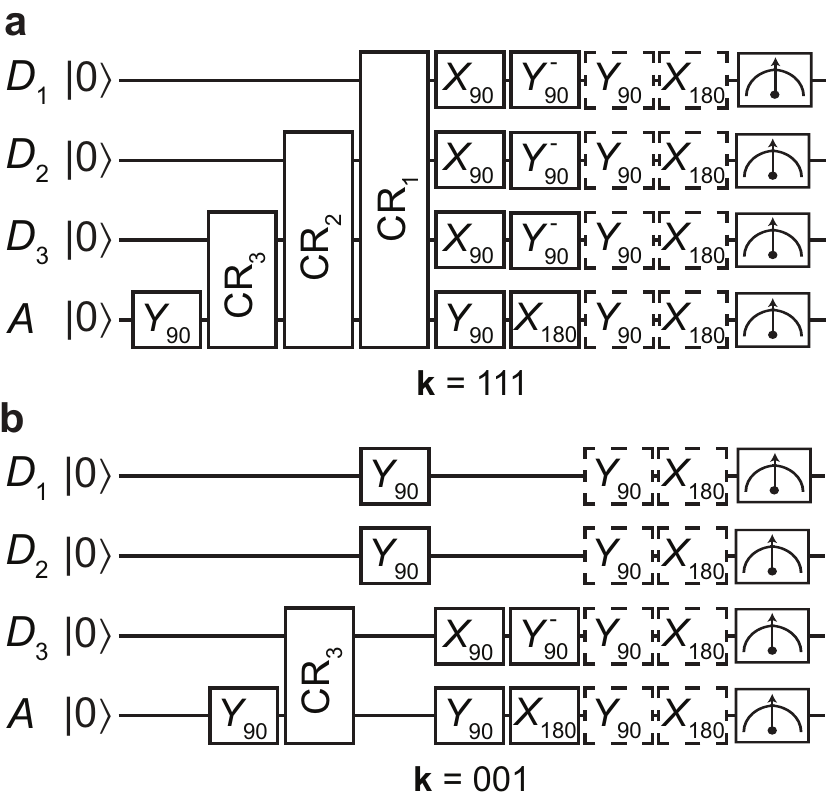}
\caption{\textbf{Circuit gate decomposition for 3-bit oracles.} \textbf{a} $\key = 111$, \textbf{b}, $\key = 001$. $\CNOT$ gates [see Fig.~1(a)] are implemented by dressing the two-qubit primitives $\ZX_i = Z_A X_{D_i}(\pi/2)$ with single-qubit gates (see Methods). Some of these gates cancel out with either state preparation (for $\Da$-$\Dc$) or with those in a subsequent $\CNOT$ gate (for $\A$) and are therefore not executed. Virtual $\Zpihalf$ gates (not shown) are applied to $\A$ after each $\ZX$ gate. Dashed boxes indicate the Hadamard decomposition applied in Q. Pulse durations are not to scale. Note that in (b) the state preparation of $\Da$ and $\Db$ is moved after $\ZX_3$ to prevent dephasing induced by the off-resonant drive.}
\end{figure*}

\begin{figure*}
\includegraphics[width=1.6\columnwidth]{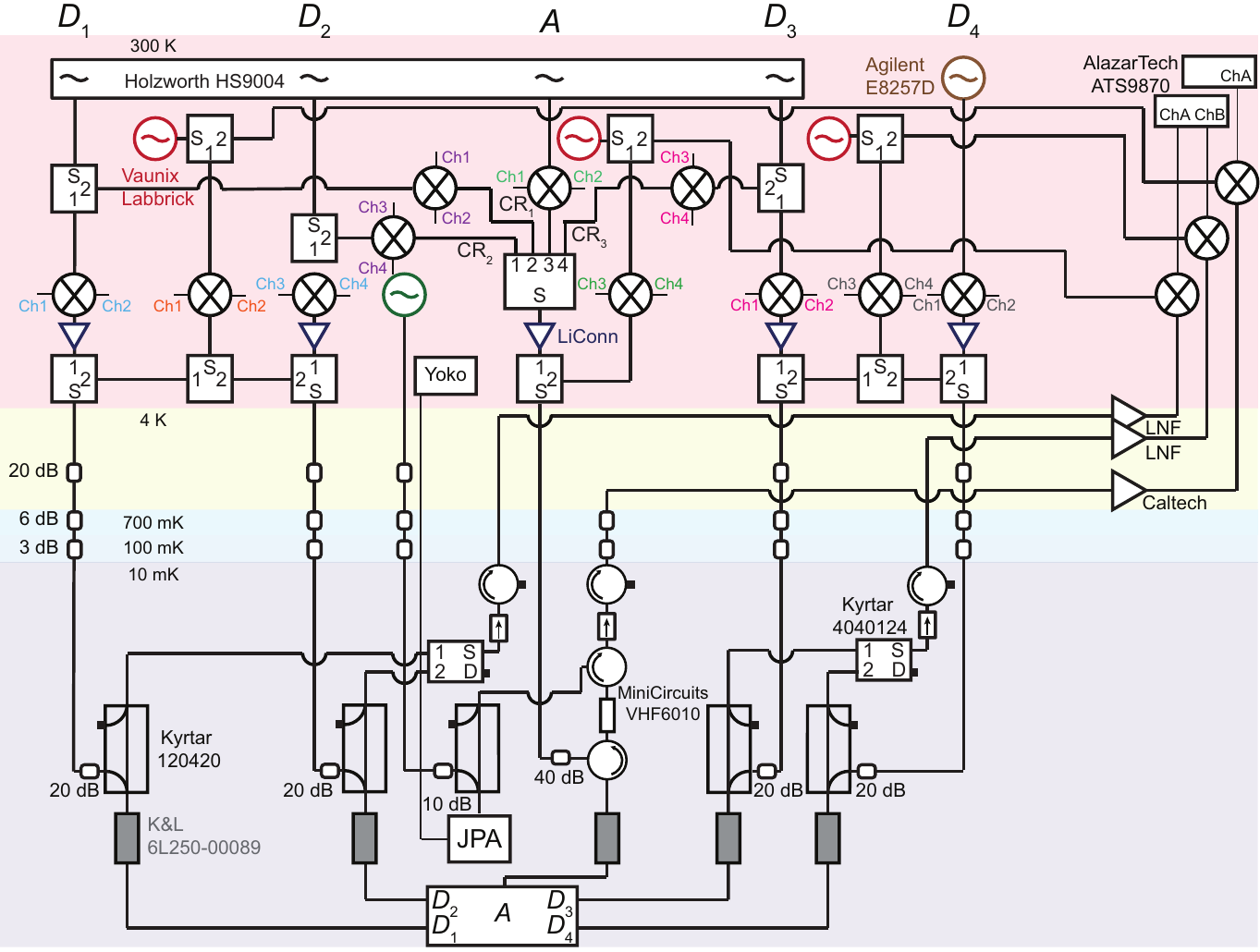}
\caption{\textbf{Experimental setup.} Complete wiring of control and readout electronics inside and outside the Bluefors BF-LD400 dilution refrigerator (see Methods). Home-made Arbitrary Pulse Sequencers (BBN APS, each indicated by its 4 analog channels Ch1-Ch4) produce the waveforms for single-qubit measurement, control, and CR pulses. The readout signal for A is boosted by a Josepshon parametric amplifier (JPA) from UC Berkeley~\cite{Hatridge11}.}
\end{figure*}

\begin{figure*}
\includegraphics[width=1.3\columnwidth]{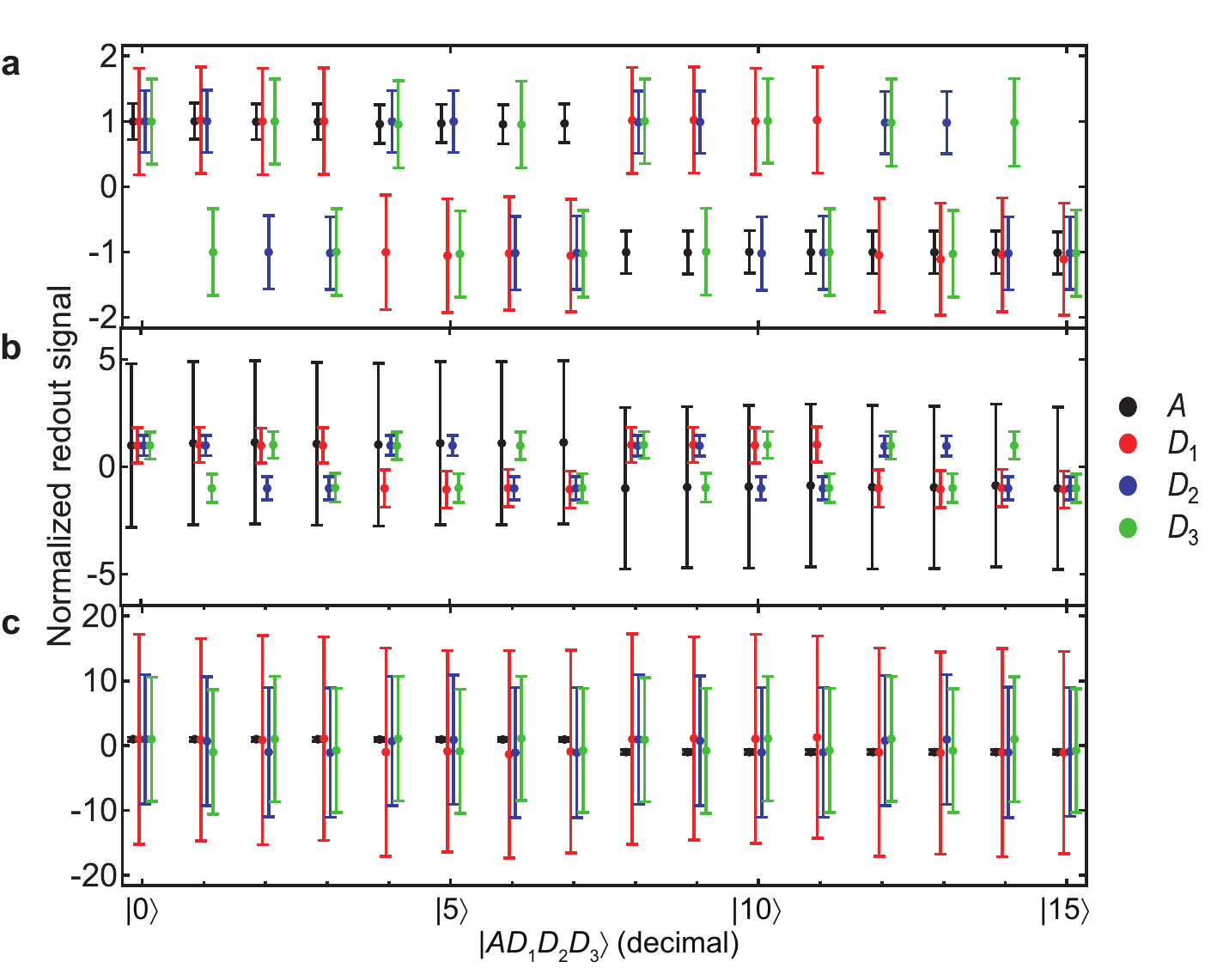}
\caption{\textbf{Readout voltage distributions.} Normalized readout signals for $\A, \Da, \Db, \Dc$ for the 16 4-qubit computational states at optimum readout settings (comparable to Figs.~2-3) (a), and for the maximum $\etaa$ (b) and $\etad$ (c) in Fig.~4a and b, respectively. Dots and error bars indicate averages and standard deviations, respectively. These data are taken in a subsequent cooldown of the same device under similar conditions, but with qubit transitions shifted up in frequency by $\sim20~\MHz$.}
\end{figure*}

\begin{figure*}
\includegraphics[width=1.5\columnwidth]{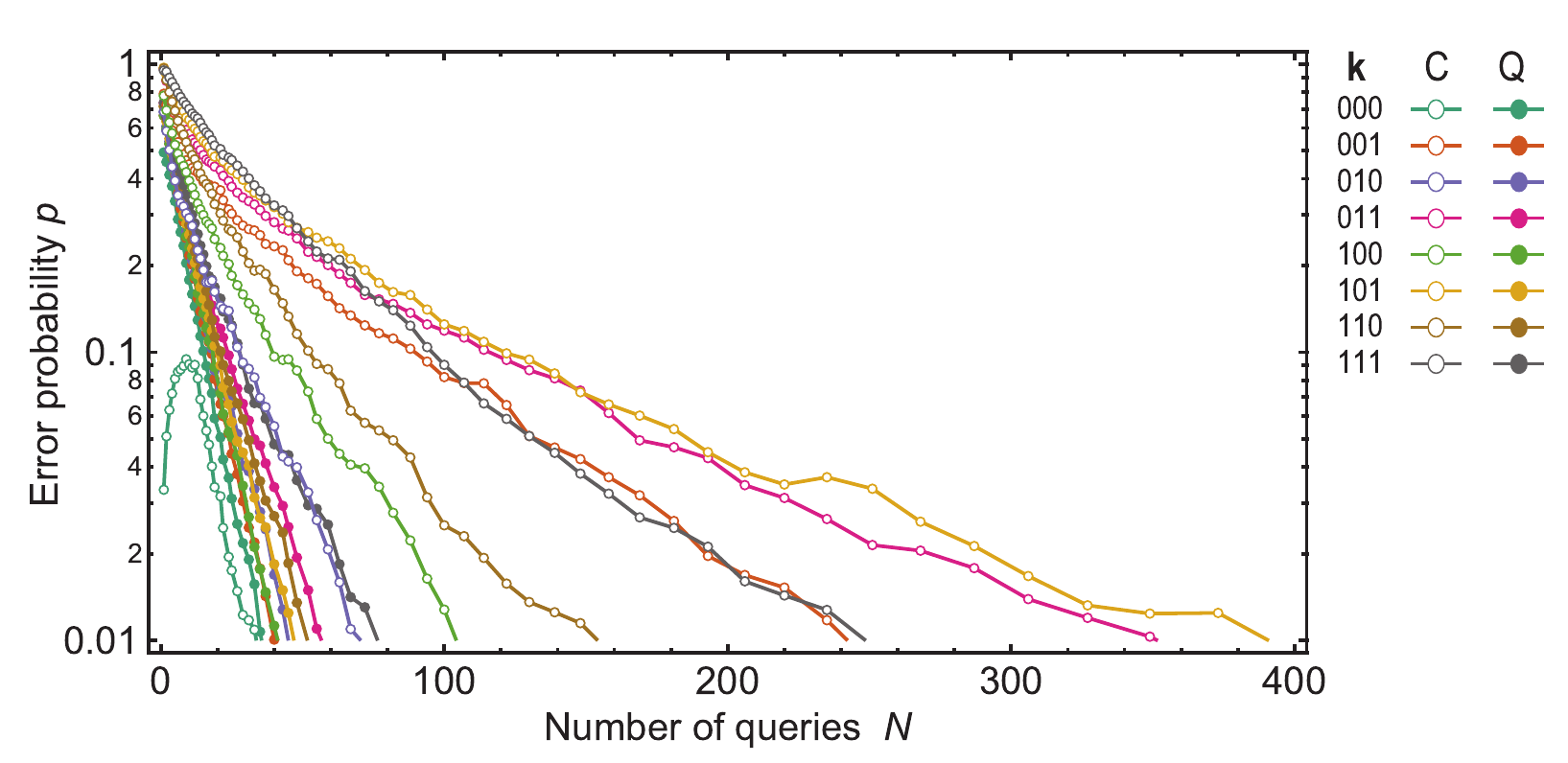}
\caption{\textbf{Learning error $\pe$ for the individual 3-bit $\key$.} The oracle queries are processed by the analog C (empty symbols) and Q (solid) solvers. The average errors are shown in Fig.~3b.}
\end{figure*}

\begin{figure*}
\includegraphics[width=1.5\columnwidth]{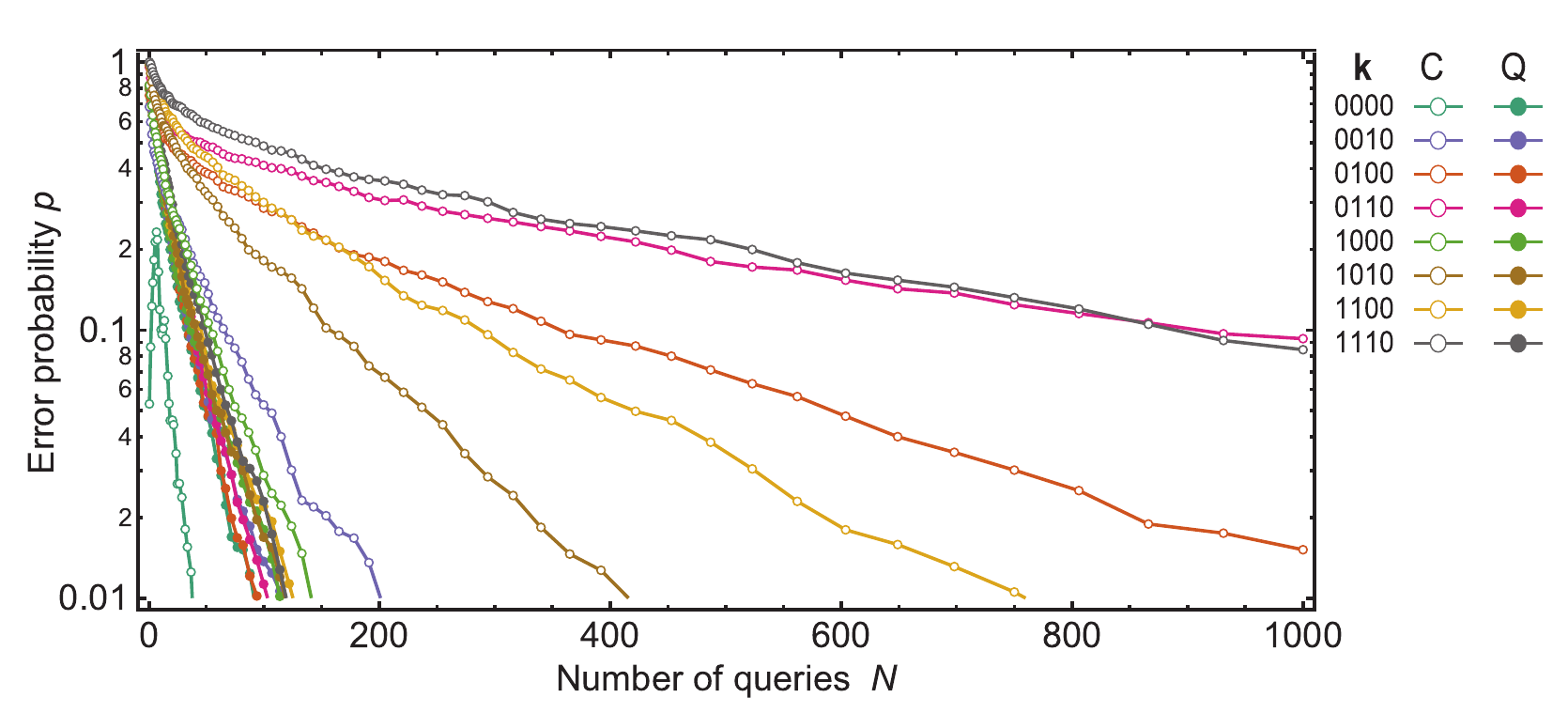}
\caption{\textbf{Learning error $\pe$ for 4-bit oracles}. Only the oracles with $k_4 = 0$ could be implemented in this device.}
\end{figure*}

\begin{figure*}
\includegraphics[width=1.3\columnwidth]{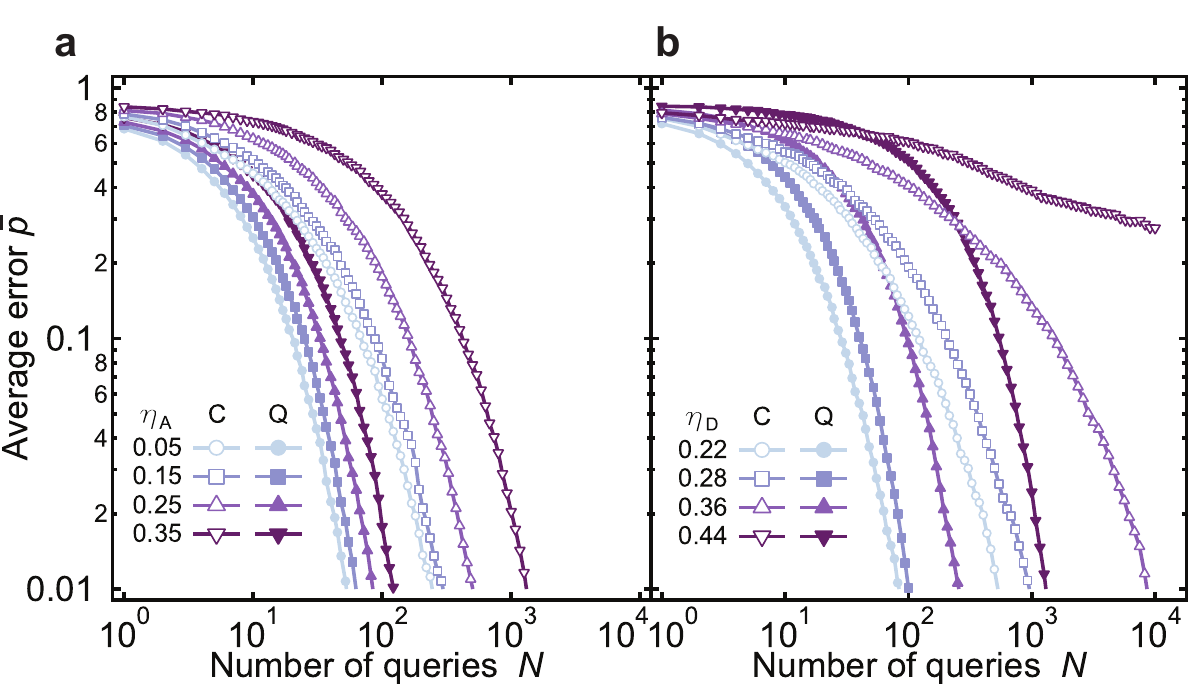}
\caption{\textbf{Average learning error $\peavg$ as a function of readout errors}. The outputs of 3-bit oracles are corrupted by increasing $\etaa$ (\textbf{a}) or $\etad$ (\textbf{b}). The intercepts of these (and additional) curves with $\peavg=0.01$ are shown in Fig.~4.}
\end{figure*}

\begin{figure*}
\includegraphics[width=0.8\columnwidth]{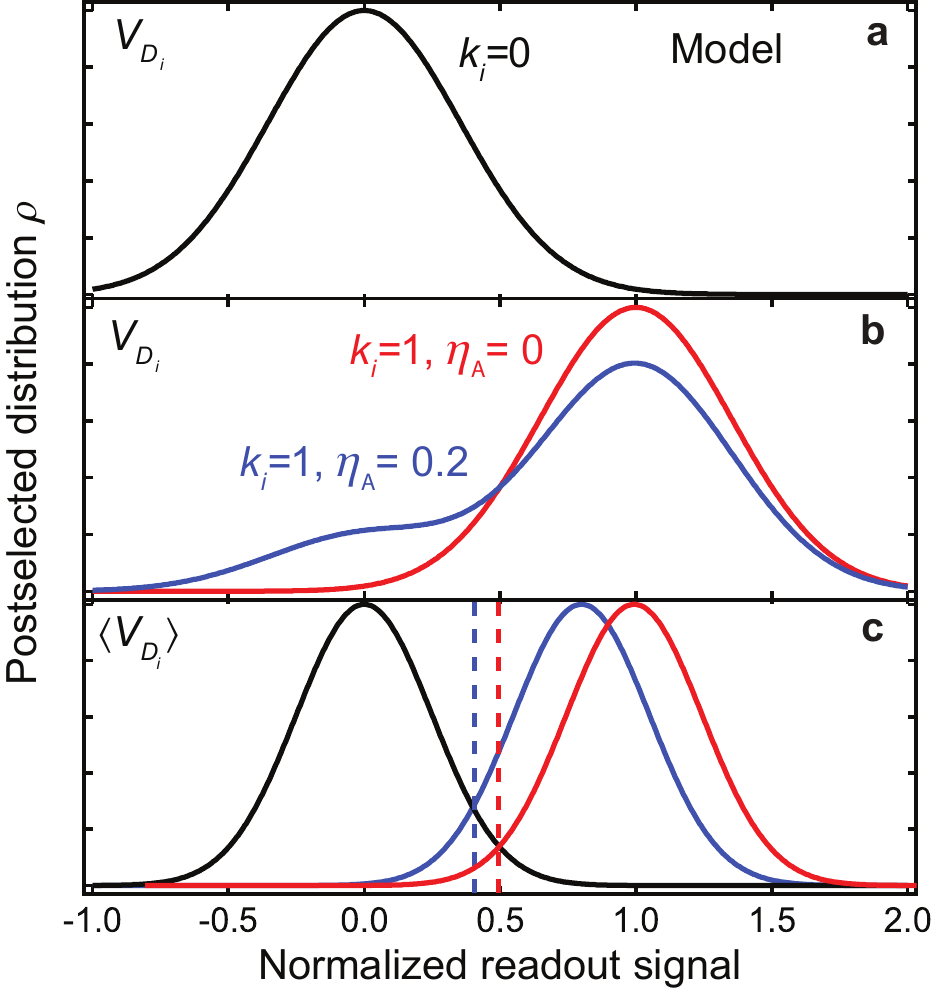}
\caption{\textbf{Calibration of the digitization threshold $\Vdi$ for the analog quantum solver Q}. For illustration purposes, we assume that the $\Pdz$ and $\Pdo$ (see Methods) have mean equal to 0 and 1, respectively, and variance equal to 0.25 in both cases. Ignoring oracle errors, $\Pkz$ (a) coincides with $\Pdz$, while $\Pko$ (b) is a mixture of the two, with weights determined by the postselection error $\etaa$, here 0 (red) or 0.2 (blue). (c) Distribution of the mean $\avg{\Vdi}$. Increasing $\etaa$ shifts the mean towards 0, thus decreasing the optimum discrimination threshold. Variances are arbitrarily scaled by a factor of 2, which does not affect the choice of threshold. The case without postselection on the ancilla ($\mathrm{Q}'$) corresponds to $\etaa = 0.5$ (not shown) for the purpose of determining the threshold.}
\end{figure*}

\renewcommand{\figurename}{Extended Data Table}
\setcounter{figure}{0}  

\begin{figure*}
\includegraphics[width=1.2\columnwidth]{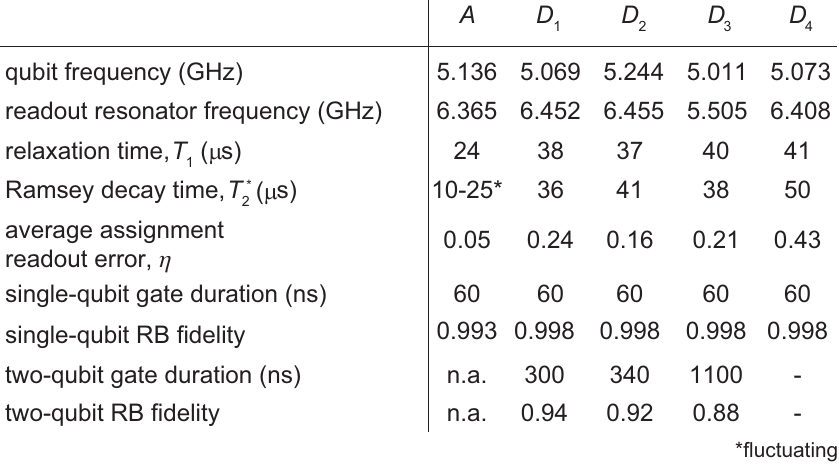}
\caption{\textbf{Qubit and resonator parameters.} Single- and two-qubit gate fidelities are obtained by randomized benchmarking (RB)~\cite{Magesan12}.}
\end{figure*}


\begin{thebibliography}{10}
\expandafter\ifx\csname url\endcsname\relax
  \def\url#1{\texttt{#1}}\fi
\expandafter\ifx\csname urlprefix\endcsname\relax\def\urlprefix{URL }\fi
\providecommand{\bibinfo}[2]{#2}
\providecommand{\eprint}[2][]{\url{#2}}

\bibitem{Nielsen00}
\bibinfo{author}{Nielsen, M.~A.} \& \bibinfo{author}{Chuang, I.~L.}
\newblock \emph{\bibinfo{title}{Quantum Computation and Quantum Information}}
  (\bibinfo{publisher}{Cambridge University Press},
  \bibinfo{address}{Cambridge}, \bibinfo{year}{2000}).

\bibitem{Cleve01}
\bibinfo{author}{Cleve, R.}
\newblock \bibinfo{title}{An introduction to quantum complexity theory}.
\newblock In \emph{\bibinfo{booktitle}{Collected Papers on Quantum Computation
  and Quantum Information Theory}}, \bibinfo{pages}{103--127}
  (\bibinfo{publisher}{World Scientific}, \bibinfo{year}{2001}).

\bibitem{Jones98}
\bibinfo{author}{Jones, J.~A.}, \bibinfo{author}{Mosca, M.} \&
  \bibinfo{author}{Hansen, R.~H.}
\newblock \bibinfo{title}{Implementation of a quantum search algorithm on a
  quantum computer}.
\newblock \emph{\bibinfo{journal}{Nature}} \textbf{\bibinfo{volume}{393}},
  \bibinfo{pages}{344--346} (\bibinfo{year}{1998}).

\bibitem{Linden98}
\bibinfo{author}{Linden, N.}, \bibinfo{author}{Barjat, H.} \&
  \bibinfo{author}{Freeman, R.}
\newblock \bibinfo{title}{An implementation of the {D}eutsch-{J}ozsa algorithm
  on a three-qubit {NMR} quantum computer}.
\newblock \emph{\bibinfo{journal}{J. Phys. Chem.}}
  \textbf{\bibinfo{volume}{296}}, \bibinfo{pages}{61 -- 67}
  (\bibinfo{year}{1998}).

\bibitem{Chuang98a}
\bibinfo{author}{Chuang, I.~L.}, \bibinfo{author}{Vandersypen, L. M.~K.},
  \bibinfo{author}{Zhou, X.}, \bibinfo{author}{Leung, D.~W.} \&
  \bibinfo{author}{Lloyd, S.}
\newblock \bibinfo{title}{Experimental realization of a quantum algorithm}.
\newblock \emph{\bibinfo{journal}{Nature}} \textbf{\bibinfo{volume}{393}},
  \bibinfo{pages}{143--146} (\bibinfo{year}{1998}).

\bibitem{Chuang98b}
\bibinfo{author}{Chuang, I.~L.}, \bibinfo{author}{Gershenfeld, N.} \&
  \bibinfo{author}{Kubinec, M.}
\newblock \bibinfo{title}{Experimental implementation of fast quantum
  searching}.
\newblock \emph{\bibinfo{journal}{Phys. Rev. Lett.}}
  \textbf{\bibinfo{volume}{80}}, \bibinfo{pages}{3408} (\bibinfo{year}{1998}).

\bibitem{Gulde03}
\bibinfo{author}{Gulde, S.} \emph{et~al.}
\newblock \bibinfo{title}{{Implementation of the {D}eutsch-{J}ozsa algorithm on
  an ion-trap quantum computer.}}
\newblock \emph{\bibinfo{journal}{Nature}} \textbf{\bibinfo{volume}{421}},
  \bibinfo{pages}{48--50} (\bibinfo{year}{2003}).

\bibitem{Takeuchi00}
\bibinfo{author}{Takeuchi, S.}
\newblock \bibinfo{title}{Experimental demonstration of a three-qubit quantum
  computation algorithm using a single photon and linear optics}.
\newblock \emph{\bibinfo{journal}{Phys. Rev. A}} \textbf{\bibinfo{volume}{62}},
  \bibinfo{pages}{032301} (\bibinfo{year}{2000}).

\bibitem{Kwiat00}
\bibinfo{author}{Kwiat, P.~G.}, \bibinfo{author}{Mitchell, J.~R.},
  \bibinfo{author}{Schwindt, P. D.~D.} \& \bibinfo{author}{White, A.~G.}
\newblock \bibinfo{title}{Grover's search algorithm: an optical approach}.
\newblock \emph{\bibinfo{journal}{J. Mod. Opt.}} \textbf{\bibinfo{volume}{47}},
  \bibinfo{pages}{257--266} (\bibinfo{year}{2000}).

\bibitem{DiCarlo09}
\bibinfo{author}{Di{C}arlo, L.} \emph{et~al.}
\newblock \bibinfo{title}{Demonstration of two-qubit algorithms with a
  superconducting quantum processor}.
\newblock \emph{\bibinfo{journal}{Nature}} \textbf{\bibinfo{volume}{460}},
  \bibinfo{pages}{240} (\bibinfo{year}{2009}).

\bibitem{Yamamoto10}
\bibinfo{author}{Yamamoto, T.} \emph{et~al.}
\newblock \bibinfo{title}{Quantum process tomography of two-qubit
  controlled-{Z} and controlled-{NOT} gates using superconducting phase
  qubits}.
\newblock \emph{\bibinfo{journal}{Phys. Rev. B}} \textbf{\bibinfo{volume}{82}},
  \bibinfo{pages}{184515} (\bibinfo{year}{2010}).

\bibitem{Dewes12}
\bibinfo{author}{Dewes, A.} \emph{et~al.}
\newblock \bibinfo{title}{{Quantum speeding-up of computation demonstrated in a
  superconducting two-qubit processor}}.
\newblock \emph{\bibinfo{journal}{Phys. Rev. B}} \textbf{\bibinfo{volume}{85}},
  \bibinfo{pages}{140503} (\bibinfo{year}{2012}).

\bibitem{Angluin88}
\bibinfo{author}{Angluin, D.} \& \bibinfo{author}{Laird, P.}
\newblock \bibinfo{title}{Learning from noisy examples}.
\newblock \emph{\bibinfo{journal}{Machine Learning}}
  \textbf{\bibinfo{volume}{2}}, \bibinfo{pages}{343--370}
  (\bibinfo{year}{1988}).

\bibitem{Blum03}
\bibinfo{author}{Blum, A.}, \bibinfo{author}{Kalai, A.} \&
  \bibinfo{author}{Wasserman, H.}
\newblock \bibinfo{title}{Noise-tolerant learning, the parity problem, and the
  statistical query model}.
\newblock \emph{\bibinfo{journal}{J. ACM}} \textbf{\bibinfo{volume}{50}},
  \bibinfo{pages}{506--519} (\bibinfo{year}{2003}).

\bibitem{Cross15}
\bibinfo{author}{Cross, A.~W.}, \bibinfo{author}{Smith, G.} \&
  \bibinfo{author}{Smolin, J.~A.}
\newblock \bibinfo{title}{Quantum learning robust against noise}.
\newblock \emph{\bibinfo{journal}{Phys. Rev. A}} \textbf{\bibinfo{volume}{92}},
  \bibinfo{pages}{012327} (\bibinfo{year}{2015}).

\bibitem{Schuld15}
\bibinfo{author}{Schuld, M.}, \bibinfo{author}{Sinayskiy, I.} \&
  \bibinfo{author}{Petruccione, F.}
\newblock \bibinfo{title}{An introduction to quantum machine learning}.
\newblock \emph{\bibinfo{journal}{Contemporary Physics}}
  \textbf{\bibinfo{volume}{56}}, \bibinfo{pages}{172--185}
  (\bibinfo{year}{2015}).

\bibitem{Manzano09}
\bibinfo{author}{Manzano, D.}, \bibinfo{author}{Pawłowski, M.} \&
  \bibinfo{author}{\ifmmode \check{C}\else~\v{C}\fi{}. Brukner}.
\newblock \bibinfo{title}{The speed of quantum and classical learning for
  performing the k th root of {NOT}}.
\newblock \emph{\bibinfo{journal}{New J. Phys.}} \textbf{\bibinfo{volume}{11}},
  \bibinfo{pages}{113018} (\bibinfo{year}{2009}).

\bibitem{Lloyd13}
\bibinfo{author}{Lloyd, S.}, \bibinfo{author}{Mohseni, M.} \&
  \bibinfo{author}{Rebentrost, P.}
\newblock \bibinfo{title}{Quantum algorithms for supervised and unsupervised
  machine learning}.
\newblock \emph{\bibinfo{journal}{arXiv:quant-ph/1307.0411}}
  (\bibinfo{year}{2013}).

\bibitem{Wiebe14}
\bibinfo{author}{Wiebe, N.}, \bibinfo{author}{Granade, C.},
  \bibinfo{author}{Ferrie, C.} \& \bibinfo{author}{Cory, D.~G.}
\newblock \bibinfo{title}{Hamiltonian learning and certification using quantum
  resources}.
\newblock \emph{\bibinfo{journal}{Phys. Rev. Lett.}}
  \textbf{\bibinfo{volume}{112}}, \bibinfo{pages}{190501}
  (\bibinfo{year}{2014}).

\bibitem{Lyubashevsky05}
\bibinfo{author}{Lyubashevsky, V.}
\newblock \emph{\bibinfo{title}{{Approximation, Randomization and Combinatorial
  Optimization. Algorithms and Techniques}}}, vol. \bibinfo{volume}{3624} of
  \emph{\bibinfo{series}{Lecture Notes in Computer Science}}
  (\bibinfo{publisher}{Springer Berlin Heidelberg}, \bibinfo{address}{Berlin,
  Heidelberg}, \bibinfo{year}{2005}).

\bibitem{Hopper01}
\bibinfo{author}{Hopper, N.} \& \bibinfo{author}{Blum, M.}
\newblock \bibinfo{title}{Secure human identification protocols}.
\newblock In \emph{\bibinfo{booktitle}{Advances in Cryptology — ASIACRYPT
  2001}}, vol. \bibinfo{volume}{2248} of \emph{\bibinfo{series}{Lecture Notes
  in Computer Science}}, \bibinfo{pages}{52--66} (\bibinfo{publisher}{Springer
  Berlin Heidelberg}, \bibinfo{year}{2001}).

\bibitem{Pietrzak12}
\bibinfo{author}{Pietrzak, K.}
\newblock \emph{\bibinfo{title}{{SOFSEM 2012: Theory and Practice of Computer
  Science}}}, vol. \bibinfo{volume}{7147} of \emph{\bibinfo{series}{Lecture
  Notes in Computer Science}} (\bibinfo{publisher}{Springer Berlin Heidelberg},
  \bibinfo{address}{Berlin, Heidelberg}, \bibinfo{year}{2012}).

\bibitem{Corcoles15}
\bibinfo{author}{C\'{o}rcoles, A.} \emph{et~al.}
\newblock \bibinfo{title}{{Demonstration of a quantum error detection code
  using a square lattice of four superconducting qubits}}.
\newblock \emph{\bibinfo{journal}{Nature Comm.}} \textbf{\bibinfo{volume}{6}},
  \bibinfo{pages}{6979} (\bibinfo{year}{2015}).

\bibitem{Blais04}
\bibinfo{author}{Blais, A.}, \bibinfo{author}{Huang, R.-S.},
  \bibinfo{author}{Wallraff, A.}, \bibinfo{author}{Girvin, S.~M.} \&
  \bibinfo{author}{Schoelkopf, R.~J.}
\newblock \bibinfo{title}{Cavity quantum electrodynamics for superconducting
  electrical circuits: An architecture for quantum computation}.
\newblock \emph{\bibinfo{journal}{Phys. Rev. A}} \textbf{\bibinfo{volume}{69}},
  \bibinfo{pages}{062320} (\bibinfo{year}{2004}).

\bibitem{Rigetti10}
\bibinfo{author}{Rigetti, C.} \& \bibinfo{author}{Devoret, M.}
\newblock \bibinfo{title}{Fully microwave-tunable universal gates in
  superconducting qubits with linear couplings and fixed transition
  frequencies}.
\newblock \emph{\bibinfo{journal}{Phys. Rev. B}} \textbf{\bibinfo{volume}{81}},
  \bibinfo{pages}{134507} (\bibinfo{year}{2010}).

\bibitem{Magesan12}
\bibinfo{author}{Magesan, E.}, \bibinfo{author}{Gambetta, J.~M.} \&
  \bibinfo{author}{Emerson, J.}
\newblock \bibinfo{title}{Characterizing quantum gates via randomized
  benchmarking}.
\newblock \emph{\bibinfo{journal}{Phys. Rev. A}} \textbf{\bibinfo{volume}{85}},
  \bibinfo{pages}{042311} (\bibinfo{year}{2012}).

\bibitem{Hatridge11}
\bibinfo{author}{Hatridge, M.}, \bibinfo{author}{Vijay, R.},
  \bibinfo{author}{Slichter, D.~H.}, \bibinfo{author}{Clarke, J.} \&
  \bibinfo{author}{Siddiqi, I.}
\newblock \bibinfo{title}{Dispersive magnetometry with a quantum limited
  {SQUID} parametric amplifier}.
\newblock \emph{\bibinfo{journal}{Phys. Rev. B}} \textbf{\bibinfo{volume}{83}},
  \bibinfo{pages}{134501} (\bibinfo{year}{2011}).

\bibitem{Vijay11}
\bibinfo{author}{Vijay, R.}, \bibinfo{author}{Slichter, D.~H.} \&
  \bibinfo{author}{Siddiqi, I.}
\newblock \bibinfo{title}{{Observation of quantum jumps in a superconducting
  artificial atom}}.
\newblock \emph{\bibinfo{journal}{Phys. Rev. Lett.}}
  \textbf{\bibinfo{volume}{106}}, \bibinfo{pages}{110502}
  (\bibinfo{year}{2011}).

\bibitem{Kimmel15}
\bibinfo{author}{Kimmel, S.}, \bibinfo{author}{Low, G.~H.} \&
  \bibinfo{author}{Yoder, T.~J.}
\newblock \bibinfo{title}{Robust calibration of a universal single-qubit
  gate-set via robust phase estimation}.
\newblock \emph{\bibinfo{journal}{arXiv:quant-ph/1502.02677}}
  (\bibinfo{year}{2015}).

\bibitem{Motzoi09}
\bibinfo{author}{Motzoi, F.}, \bibinfo{author}{Gambetta, J.~M.},
  \bibinfo{author}{Rebentrost, P.} \& \bibinfo{author}{Wilhelm, F.~K.}
\newblock \bibinfo{title}{Simple pulses for elimination of leakage in weakly
  nonlinear qubits}.
\newblock \emph{\bibinfo{journal}{Phys. Rev. Lett.}}
  \textbf{\bibinfo{volume}{103}}, \bibinfo{pages}{110501}
  (\bibinfo{year}{2009}).

\bibitem{Lucero10}
\bibinfo{author}{Lucero, E.} \emph{et~al.}
\newblock \bibinfo{title}{Reduced phase error through optimized control of a
  superconducting qubit}.
\newblock \emph{\bibinfo{journal}{Phys. Rev. A}} \textbf{\bibinfo{volume}{82}},
  \bibinfo{pages}{042339} (\bibinfo{year}{2010}).

\bibitem{Jerger12}
\bibinfo{author}{Jerger, M.} \emph{et~al.}
\newblock \bibinfo{title}{Frequency division multiplexing readout and
  simultaneous manipulation of an array of flux qubits}.
\newblock \emph{\bibinfo{journal}{Appl. Phys. Lett.}}
  \textbf{\bibinfo{volume}{101}}, \bibinfo{pages}{042604}
  (\bibinfo{year}{2012}).

\bibitem{Chow14}
\bibinfo{author}{Chow, J.~M.} \emph{et~al.}
\newblock \bibinfo{title}{{Implementing a strand of a scalable fault-tolerant
  quantum computing fabric.}}
\newblock \emph{\bibinfo{journal}{Nature Comm.}} \textbf{\bibinfo{volume}{5}},
  \bibinfo{pages}{4015} (\bibinfo{year}{2014}).

\end{thebibliography}
\end{document}